%% file: eprint.tex
\documentclass[12pt]{article}
\usepackage{graphicx}

\newcommand\pubnumber{SNSN-323-63}
\newcommand\pubdate{\today}

\def\Title#1{\begin{center} {\Large #1 } \end{center}}
\def\Author#1{\begin{center}{ \sc #1} \end{center}}
\def\Address#1{\begin{center}{ \it #1} \end{center}}

\newcommand\pubblock{\rightline{\begin{tabular}{l} \pubnumber\\
         \pubdate  \end{tabular}}}
\newenvironment{Abstract}{\begin{quotation}  }{\end{quotation}}
\newenvironment{Presented}{\begin{quotation} \begin{center} 
             PRESENTED AT\end{center}\bigskip 
      \begin{center}\begin{large}}{\end{large}\end{center} \end{quotation}}
\def\Acknowledgements{\bigskip  \bigskip \begin{center} \begin{large}
             \bf ACKNOWLEDGEMENTS \end{large}\end{center}}

\input econfmacros.tex

\begin{document}
\begin{titlepage}
\pubblock

\vfill
\Title{Non Perturbative One Gluon Exchange Potential from
Dyson-Schwinger Equations}
\vfill
\Author{P. Gonz\'{a}lez$^1$, V. Vento$^1$}
\Address{$^1$Departamento de F\'{\i}sica
Te\'{o}rica -IFIC. Universidad de Valencia-CSIC. E-46100
Burjassot (Valencia), Spain.}
\Author{V. Mathieu$^2$}
\Address{$^2$ECT$^{\ast}$, European Center for
Theoretical Studies in Nuclear Physics and Related Areas, Strada
delle Tabarelle 286, I-38123 Villazzano (TN), Italy}
\vfill
\begin{Abstract}
Recent progress in the solution of
Dyson-Schwinger equations of QCD allows for a non perturbative evaluation of
the One Gluon Exchange (OGE) interaction. We calculate the interquark static
potential for heavy mesons by assuming that it is given by this OGE
interaction and we apply it to the description of charmonium.
\end{Abstract}
\vfill
\begin{Presented}
Charm 2012. The 5$^{th}$ International Workshop \\on Charm Physics. Hawaii, USA.
\end{Presented}
\vfill
\end{titlepage}
\setcounter{footnote}{0}

\section{Introduction}

The development of non-perturbative techniques in Quantum Chromodynamics (QCD)
is starting to allow for a description of the hadron spectrum from first
principles. Among these techniques Lattice gauge theory
\cite{Wilson:1974sk,Creutz:1980zw} constitutes a non-perturbative
regularisation scheme which may provide accurate numerical solutions of the
theory. The accuracy of lattice results has been tremendously improved during
the past decade with the availability of more powerful computers
\cite{Sachrajda:2011tg,Hagler:2011zz}. Therefore, lattice results are
considered in many instances the data which other non perturbative
approximations try to reproduce. One of these alternatives is the approximate
resolution of the Dyson-Schwinger Equations (DSE) of QCD, a non-perturbative
approach which has progressed considerably in the last ten years. This
approach, more analytical, has led to a very appealing physical picture
establishing that the QCD running coupling (effective color charge) freezes in
the deep infrared. This property can be best understood from the point of view
of a dynamical gluon mass generation that is a purely non-perturbative effect
\cite{Cornwall:1982zr,Aguilar:2006gr,Binosi:2009qm}.

The aim of this presentation is to investigate, following reference
\cite{Gon11}, the form of the OGE static potential from DSE and compare it to
the static potentials derived from lattice calculations. The application of
these potentials to the description of quarkonia will be discussed.

\section{One Gluon Exchange Potential from Dyson \\Schwinger Equations}

It is well established by now that the QCD running coupling (effective charge)
freezes in the deep infrared what may be understood in terms of a dynamical
gluon mass. Even though the gluon is massless at the level of the fundamental
QCD Lagrangian, and remains massless to all order in perturbation theory, the
non-perturbative QCD dynamics generate an effective, momentum-dependent mass,
without affecting the local $SU(3)_{c}$ invariance, which remains intact
\cite{Cornwall:1982zr,Aguilar:2006gr,Binosi:2009qm,Aguilar:2008xm}. At the
level of the Dyson-Schwinger equations (DSE), solved by using a PT (pinch
technique) - BFM (background field method) truncation scheme in the quenched
approximation (no quark loops), the generation of such a mass is associated
with the existence of infrared finite solutions for the gluon propagator. Such
solutions may be fitted by \textquotedblleft massive\textquotedblright%
\ euclidean propagators of the form $\Delta^{-1}(q^{2})=q^{2}+m^{2}(q^{2})$
where $m^{2}(q^{2})$ depends non-trivially on the momentum transfer $q^{2}$.

One physically motivated possibility is the so called logarithmic mass
running, which is defined by%

\begin{equation}
m^{2}(q^{2})=m_{0}^{2}\left[  \ln\left(  \frac{q^{2}+\rho m_{0}^{2}}%
{\Lambda^{2}}\right)  \bigg/\ln\left(  \frac{\rho m_{0}^{2}}{\Lambda^{2}%
}\right)  \right]  ^{-1-\delta}. \label{rmass}%
\end{equation}
where $\Lambda$ is the QCD$\ $scale and $m_{0},$ the gluon mass at
$q^{2}\rightarrow0,$ as well as $\rho$ and $\delta$ are constants to be
fitted. For this fitting the calculated DSE propagator is compared to the one
obtained from lattice.

On the other hand the non-perturbative generalization of $\alpha(q^{2}),$ the
QCD running coupling, comes in the form
\begin{equation}
\alpha(q^{2})=4\pi\left[  \beta_{0}\ln\left(  \frac{q^{2}+\rho m^{2}(q^{2}%
)}{\Lambda^{2}}\right)  \right]  ^{-1}, \label{alphalog}%
\end{equation}
where $\beta_{0}=11-2n_{f}/3$ being $n_{f}$ the number of active quark flavours.

Note that its zero gluon mass limit leads to the LO perturbative coupling
constant momentum dependence. The $m(q^{2})$ in the argument of the logarithm
tames the Landau pole, and $\alpha(q^{2})$ freezes at a finite value in the
IR, namely $\alpha(0)=4\pi\left[  \beta_{0}\ln\left(  \frac{\rho m_{0}^{2}%
}{\Lambda^{2}}\right)  \right]  ^{-1}$.

\begin{figure}[tbh]
\begin{center}
\includegraphics[width=10cm]{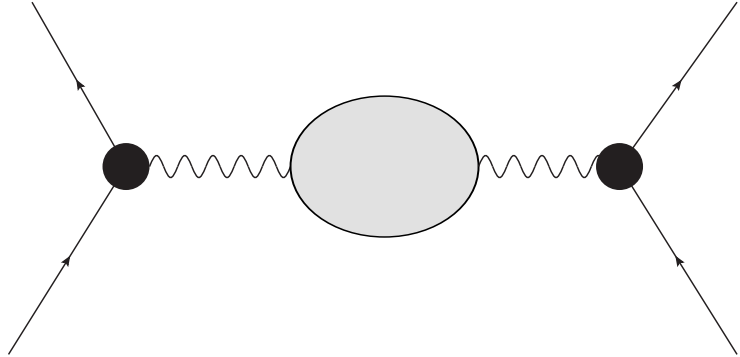}
\end{center}
\caption{One gluon exchange interaction.}%
\label{oge}%
\end{figure}

From these expressions for the propagator and the coupling we can derive the
One Gluon Exchange interaction between static charges, see Fig \ref{oge}.

The OGE static potential is related to the Fourier transform of the time-time
component of the full gluon propagator in the following way%

\begin{equation}
V({\mathbf{r}})=-C_{F}\int\frac{d^{3}{\mathbf{k}}}{(2\pi)^{3}}\,\frac
{4\pi\alpha({\mathbf{k}}^{2})}{{\mathbf{k}}^{2}+m^{2}({\mathbf{k}}^{2}%
)}e^{i{\mathbf{k.r}}}=-\frac{32\pi C_{F}}{{\mathbf{r}}}\int_{0}^{\infty
}\!\!\!d{\mathbf{k}}\;{\mathbf{k}}\frac{\alpha({\mathbf{k}}^{2})}{{\mathbf{k}%
}^{2}+m^{2}({\mathbf{k}}^{2})}\sin({{\mathbf{k}r}}) \label{pot}%
\end{equation}
where $C_{F}$ is the Casimir eigenvalue of the fundamental representation of
$SU(3)$ [$C_{F}=(N^{2}-1)/2N$ for $SU(N)$] and the bold terms, ${\mathbf{k}}$
and ${\mathbf{r}}$ stand for 3-vectors.

\begin{figure}[tbh]
\begin{center}
\includegraphics[width=10cm]{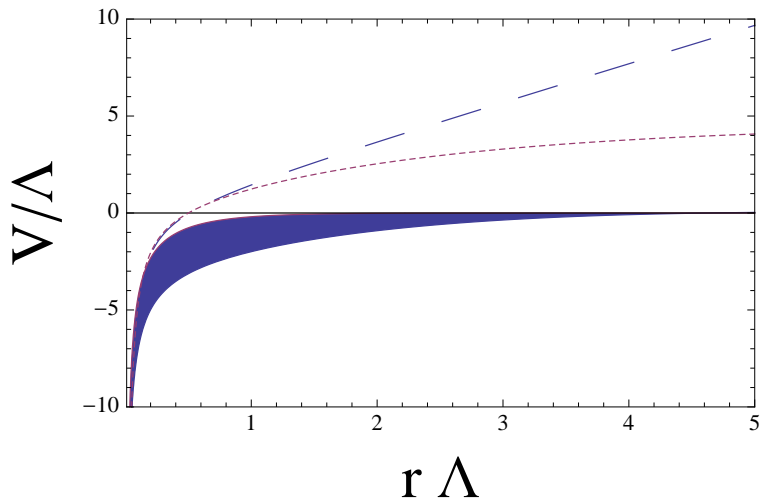}
\end{center}
\caption{The DSE OGE potential, i.e. the Fourier transform of the massive One
Gluon Exchange, is plotted for the range $m\sim360-480$ MeV and $\rho\sim1-4$
with $n_{f}=4$ ($\beta_{0}=25/3$). For comparison we plot the Cornell and
screened potentials with parameters $a=0.52,$ $\sqrt{b}=427$ MeV and
$\gamma=0.38$ fm$^{-1}.$}%
\label{DSE1pot}%
\end{figure}

In Fig. \ref{DSE1pot} we show the finite part (up to a constant) of the
potential, Eq \ref{pot}, derived from the DSE with the definitions in Eqs
\ref{alphalog} and \ref{rmass}. We have chosen the following range of
parameters: $m_{0}\sim360-480$ MeV, $\rho=1.-4.$, $\delta=1./11$
\cite{Cornwall:1982zr,Aguilar:2007ie,Aguilar:2009nf} which provide a good fit
to the lattice propagator. The value of $\Lambda$ here has been taken to be
$300$ MeV. In order to adjust the behavior at the origin to the data we have
used $\beta_{0}$ corresponding to $n_{f}=4$ flavors. To do this appropriately
one should introduce the running of the quark masses, which are at present not
well known. However, since asymptotically the masses run to zero, our way of
proceeding achieves the correct (perturbative) strength of the potential at
low $r$. The potential describes well the low radial behavior, by
construction, flattens at large $r$ going asymptotically to zero and never
becomes positive.

For comparison we have used i) a Cornell type potential (whose form is derived
from quenched lattice calculations \cite{Bali:2000gf,Greensite:2003xf})%
\begin{equation}
V(r)=-a/r+br.
\end{equation}
containing the perturbative expectation plus an additional linear term and ii)
a screened type potential (whose form was derived long time ago from an
unquenched lattice calculation \cite{Born:1989iv})
\begin{equation}
V(r)=(-a/r+br)\,\left(  \frac{1-e^{-\gamma r}}{\gamma r}\right)
\end{equation}
The values used for the parameters are based on spectroscopy (see next Section).

\begin{figure}[tbh]
\begin{center}
\includegraphics[width=10cm]{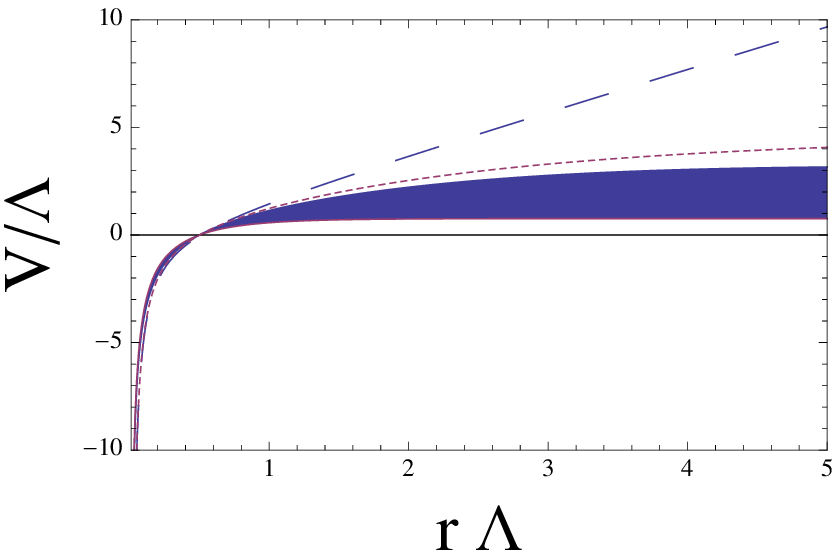}
\end{center}
\caption{The DSE OGE potential after the Sommer substraction, with the same
parameters as in the previous figure, is plotted. For comparison we plot the
Cornell and screened potentials.}%
\label{DSE1potsom}%
\end{figure}

We should realize that the additive infinite self-energy contribution
associated with the static sources should be removed from the calculated OGE
potential \cite{Bali:2000gf} . In lattice QCD this is done normalizing the
potential such that $V(r_{0})=0$ where $r_{0}$ is the Sommer scale
\cite{Sommer:1993ce} with a typical value around $0.5$ fm. We proceed the same
way but for phenomenological purposes we take the substraction point at the
zero point of the potentials used in Table 1, which happens to be at
$r_{0}\sim0.35fm$ for the parametrizations we are using. The result of this
procedure is shown in Fig. \ref{DSE1potsom}. This procedure increases the
value of the potential without changing its shape.

As can be checked our OGE potential resembles the screened and not the Cornell
potential. We have shown that none of the parameters plays a fundamental role
in the structure of the DSE OGE potential, since the structure does not vary
when we vary them. All values discussed lead to the same qualitative features
for the potential. It is evident that there is no way to reproduce the large
$r$ behavior of the Cornell potential by changing the parameters. The DSE OGE
potential flattens and becomes basically constant similar to the screened
potential. If we assume that the OGE interaction is the main source of the
dynamics, this result is quite surprising since the approximations used to
find the solution to the DSE do not contain quark loops and therefore they
incorporate no mechanism for screening, i.e. some mechanism derived from the
breaking of the string \cite{Born:1989iv,Kratochvila:2003zj}.

Therefore we should conclude that either some correction to the calculated OGE
is lacking (this could come from the use of a higher order truncation scheme
for the resolution of the DSE, from vertex corrections to the coupling ...) or
that the incorporation of multigluon effects is essential for deriving a
consistent large $r$ behavior. In this respect an \textit{ad hoc} explanation
based on OGE dominance plus additional coupling corrections has been recently
proposed along the following arguments \cite{Vento2012}. Let us consider that
there were a nonperturbative vertex correction to the strong effective charge
in the form%
\[
\alpha_{conf}(q^{2})=\frac{c\Lambda^{4}}{q^{4}}%
\]
with $c$ a constant to be fitted, so that%
\[
\alpha_{total}(q^{2})=\alpha_{conf}(q^{2})+\alpha(q^{2})
\]
where the last term on the right hand side would be given by Eq \ref{alphalog}%
. Then it would be possible to fit not only the lattice propagator but also
the Cornell type lattice potential and a fully consistent scheme in the
quenched approximation would arise. To go further one should incorporate quark
loop effects. According to reference \cite{Cor91} the former vertex correction
could be effectively modified. If this modification were parametrized through
a cutof $\left(  s\right)  $ in the form
\[
\left(  \alpha_{conf}(q^{2})\right)  _{unquenched}=\frac{d\Lambda^{4}}{\left(
q^{2}+s^{2}\right)  ^{2}}%
\]
being $d$ a constant, the resulting DSE OGE potential would be like the
screened potential. In this manner an explanation of our previous results
compatible with one gluon exchange dynamical dominance could come out. Next we
shall adopt this interpretation in the phenomenological application of the
SDSE OGE potential to charmonium.

\section{Quarkonia description}

For sufficiently heavy quarks, one may hope that the bound state problem
becomes essentially non-relativistic the dynamics being controlled
approximately by a Schr\"{o}dinger equation with a static potential. It should
be emphasized that the derived potentials do not contain spin-dependent terms
what make them more reliable when these terms do not play a major role. This
is expected to occur for high excited (large-sized) states since spin
corrections are short-ranged. For the low lying states data show that
spin-spin splittings between spin triplet and spin singlet states should be
relevant ($m_{J/\psi}-m_{\eta_{c}}=117$ MeV, $m_{\psi_{2s}}-m_{\eta_{c}%
(2s)}=49$ MeV, $m_{\Upsilon(1s)}-m_{\eta_{b}}=69$ MeV, ...). By considering
that the perturbative spin-spin correction is in absolute value three times
bigger for singlets than for triplets, the radial potential approach could be
taken as an approximate description of spin triplet states. Other
spin-dependent corrections (spin-orbit, tensor) may be playing some role. It
is worth to point out that perturbative spin-orbit and tensor splittings
cancel in the centroids of $p$ waves which consequently may be used as
\textquotedblleft data\textquotedblright\ for comparison with the radial
approach results. Further relativistic corrections are expected to be more
important for charmonium than for bottomonium. Therefore the radial approaches
are better suited for the study of bottomonium. One should not forget though
that in the application to quarkonia the parameters entering the expressions
of the potentials have an effective character since their values may be
implicitly incorporating non considered corrections.

The charmonium spectrum with well established quantum numbers, corresponding
mostly to $J^{PC}=1^{--}$ resonances produced through ISR (Initial State
Radiation) processes, is analized in Table 1. The calculated spectrum from the
DSE OGE potential is compared to experimental data and to a Cornell like
potential calculation (the choice of charmonium instead of bottomonium makes
clearer the differences between both calculations). The parameters for the
Cornell potential have been chosen within the conventional spectroscopic range
$a\sim0.51-0.52$ and $\sqrt{b}\sim412-427$ MeV
\cite{Quigg:1979vr,Eichten:1979ms} (note that concerning the description of
quarkonia masses the addditive constant in the potential can be absorbed in a
renormalization of the quark mass).

\begin{table}[tbh]
\begin{center}
\begin{tabular}
[c]{|cccc|}\hline
$n_{r}L$ & $M_{Cornell}$ & $M_{DSE}$ & $M_{PDG}$\\\hline
& MeV & MeV & MeV\\\hline
$1s$ & 3069 & 3151 & $3096.916\pm0.011$\\
$2s$ & 3688 & 3660 & $3686.09\pm0.04$\\
$1d$ & 3806 & 3761 & $3772.92\pm0.35$\\
$3s$ & 4147 & 4004 & $4039\pm1$\\
$2d$ & 4228 & 4070 & $4153\pm3$\\
$4s$ & 4539 & 4273 & $4263_{-9}^{+8}$\\
$3d$ & 4601 & 4321 & $4361\pm9\pm9$\\
$5s$ & 4829 & 4487 & $4421\pm4$\\
$4d$ & 4879 & 4526 & \\
$6s$ & 5218 & 4651 & $4664\pm11\pm5$\\\hline\hline
$1p$ & 3502 & 3515 & $3525.3\pm0.2$\\
$2p$ & 3983 & 3886 & \\\hline
\end{tabular}
\end{center}
\caption{Calculated masses, $M_{Cornell}$ and $M_{DSE},$ from the Cornell and
DSE OGE potentials. For the Cornell potential $a=0.52$, $\sqrt{b}=412$ MeV and
$m_{c}=1350$ MeV. For the DSE potential $m_{0}=345.7$ MeV, $\rho=1$ and
$m_{c}=1400$ MeV. Masses for experimental candidates, $M_{PDG},$ have been
taken from \cite{PDG10}. For $p$ waves we quote the centroid of $np_{0}$,
$np_{1}$ and $np_{2}$ states. }%
\end{table}

As can be checked the main difference between the two models refers to the
description of the higher excited states. The DSE model allows in the overall
for a reasonable one to one assignment of calculated states to data (within 60
MeV difference) whereas the Cornell model, providing a good fit for the lower
states (at most 30 MeV difference with data), can not accomodate all the known
higher energy resonances but only some of them. For instance $\psi(4040),$
$\psi(4160)$ and $\psi(4415)$ may be assigned to the Cornell $3s,$ $2d$ and
$4s$ states respectively. Then other two resonances, cataloged in the Particle
Data Group Review \cite{PDG10} as $X(4260)$ and $X(4360),$ can not be
reproduced (this has motivated alternative, non $c\overline{c},$ explanations
for these states even though their properties might be understood as
corresponding to $c\overline{c}$ states, see \cite{Eich08} and references therein).

Although more complete analyses are needed before extracting any definite
conclusion these results seem to point out that the nonperturbative OGE
potential might provide a well founded approach to heavy meson spectroscopy.

\section{Summary}

We have calculated the OGE static potential from an approximate solution of
the quenched Dyson-Schwinger equations for the gluon propagator. The low $r$
behavior is determined by the well know asymptotic behavior. The large $r$
behavior is certainly non perturbative. The Sommer procedure, to avoid self
energy effects of the static charges, leads to a potential which is not
negative everywhere. The DSE with the Sommer normalization is quite similar to
a screened potential form derived from unquenched lattice calculations.

This suggests that non considered nonperturbative vertex corrections to the
strong effective coupling could be responsible for the linear confining. These
corrections might be canceled in the unquenched DSE solution providing an ad
hoc explanation of the OGE results obtained.

Taking for granted this explanation and assuming that the OGE interaction is
the main source of the dynamics we have proceeded to a calculation of the
charmonium spectrum. We have found a one to one correspondence between the
calculated states and the experimental resonances. This makes us tentatively
conclude that the nonperturbative OGE interaction may provide a significant
improvement in the description of heavy quarkonia as compared to conventional
potentials based on confinement plus perturbative OGE terms.

\Acknowledgements

This work has been supported by Ministerio de Econom\'{\i}a y Competitividad
(Spain) and EU FEDER grant FPA2010-21750-C02-01, by AIC10-D-000598, by
Consolider Ingenio 2010 CPAN (CSD2007-00042), by GVPrometeo2009/129 and by
European Integrated Infrastructure Initiative HadronPhysics3 (Grant 283286).

\end{document}

%% file: econfmacros.tex
\def\beq{\begin{equation}}
\def\eeq#1{\label{#1}\end{equation}}
\def\eeqn{\end{equation}}

\def\beqa{\begin{eqnarray}}
\def\eeqa#1{\label{#1}\end{eqnarray}}
\def\eeqan{\end{eqnarray}}

\let\bar=\overbar

\def\Dslash{\not{\hbox{\kern-4pt $D$}}}
\def\dslash{\not{\hbox{\kern-2pt $\del$}}}

\def\msb{{\bar{\ssstyle M \kern -1pt S}}}




%% file: eprint.bbl
\begin{thebibliography}{99}



\bibitem {Wilson:1974sk}K.~G.~Wilson,
Phys.\ Rev.\ D \textbf{10,} 2445 (1974).




\bibitem {Creutz:1980zw}M.~Creutz,
Phys.\ Rev.\ D \textbf{21,} 2308 (1980).
;
arXiv:1103.3304 [hep-lat].




\bibitem {Sachrajda:2011tg}C.~Sachrajda,
PoS \textbf{LATTICE2010} (2010) 018 [arXiv:1103.5959 [hep-lat]].




\bibitem {Hagler:2011zz}P.~Hagler,
Prog.\ Theor.\ Phys.\ Suppl.\ \textbf{187,} 221 (2011).




\bibitem {Cornwall:1982zr}J.~M.~Cornwall,
Phys.\ Rev.\ D \textbf{26}, 1453 (1982).




\bibitem {Aguilar:2006gr}A.~C.~Aguilar and J.~Papavassiliou,
JHEP \textbf{0612}, 012 (2006)




\bibitem {Binosi:2009qm}D.~Binosi and J.~Papavassiliou,
Phys.\ Rept.\ \textbf{479}, 1 (2009) [arXiv:0909.2536 [hep-ph]].


\bibitem {Gon11}P. Gonz\'{a}lez, V. Mathieu and V. Vento, Phys.\ Rev.\ D
\textbf{84}, 114008 (2011).



\bibitem {Aguilar:2008xm}A.~C.~Aguilar, D.~Binosi and J.~Papavassiliou,
Phys.\ Rev.\ D \textbf{78}, 025010 (2008).




\bibitem {Aguilar:2007ie}A.~C.~Aguilar and J.~Papavassiliou,
Eur.\ Phys.\ J.\ A \textbf{35}, 189 (2008).




\bibitem {Aguilar:2009nf}A.~C.~Aguilar, D.~Binosi, J.~Papavassiliou and
J.~Rodriguez-Quintero,
Phys.\ Rev.\ D \textbf{80}, 085018 (2009);




\bibitem {Bali:2000gf}G.~S.~Bali,
Phys.\ Rept.\ \textbf{343, }1-136 (2001). [hep-ph/0001312].



\bibitem {Greensite:2003xf}J.~Greensite, S.~Olejnik,
Phys.\ Rev.\ \textbf{D67} 094503 (2003). [hep-lat/0302018].



\bibitem {Born:1989iv}K.~D.~Born, E.~Laermann, N.~Pirch, T.~F.~Walsh,
P.~M.~Zerwas,
Phys.\ Rev.\ \textbf{D40 }1653-1663 (1989).



\bibitem {Sommer:1993ce}R.~Sommer,
Nucl.\ Phys.\ B \textbf{411, }839 (1994) [arXiv:hep-lat/9310022].




\bibitem {Kratochvila:2003zj}S.~Kratochvila, P.~de Forcrand,
Nucl.\ Phys.\ \textbf{B671, }103-132 (2003). [hep-lat/0306011].

\bibitem {Vento2012}V.~Vento, arXiv: 1205.2002 [hep-ph].

\bibitem {Cor91}J.~Papavassiliou and J.~M.~Cornwall, Phys.\ Rev.\ \textbf{D44}
1285 (1991).

\bibitem {PDG10}K.~Nakamura \textit{et al.} [Particle Data Group], J. Phys. G
\textbf{37}, 075201 (2010).



\bibitem {Quigg:1979vr}C.~Quigg, J.~L.~Rosner,
Phys.\ Rept.\ \textbf{56}, 167-235 (1979).



\bibitem {Eichten:1979ms}E.~Eichten, K.~Gottfried, T.~Kinoshita, K.~D.~Lane,
T.~-M.~Yan,
Phys.\ Rev.\ \textbf{D21,} 203 (1980).

\bibitem {Eich08}E. Eichten, S. Godfrey, H. Mahlke and J. Rosner, Rev. Mod.
Phys. 80, 1161 (2008).



\end{thebibliography}
